\documentclass[runningheads]{llncs}
\usepackage{graphicx}
\usepackage[table]{xcolor}
\usepackage{amsmath}
\usepackage{graphicx}
\usepackage{mwe}
\usepackage[caption=false]{subfig}
\usepackage{gensymb}
\usepackage{wrapfig}
\usepackage{diagbox}
\usepackage[export]{adjustbox}
\usepackage{todonotes}
\newsavebox{\measurebox}
\usepackage{textcomp,gensymb}
\usepackage{float}

\begin{document}
\title{Investigating the Role of Pedestrian Groups in Shared Spaces through Simulation Modeling \thanks{This paper has been published with copyright by Springer, accessible at \url{https://link.springer.com/book/10.1007/978-3-030-45718-1}}}
\titlerunning{Investigating the Role of Pedestrian Groups in Shared Spaces}
%
\author{Suhair Ahmed\thanks{S. Ahmed and F. Johora contributed equally to this work.} \and
Fatema T. Johora \and
J\"org P. M\"uller}

\authorrunning{S. Ahmed et al.}
%
\institute{Technische Universit\"at Clausthal, Clausthal-Zellerfeld, Germany \\
\email{\{suhair.ahmed,fatema.tuj.johora,joerg.mueller\}@tu-clausthal.de}}
\maketitle              
\begin{abstract}
In shared space environments, urban space is shared among different types of road users, who frequently interact with each other to negotiate priority and coordinate their trajectories. Instead of traffic rules, interactions among them are conducted by informal rules like speed limitations and by social protocols e.g., courtesy behavior.  
Social groups (socially related road users who walk together) are an essential phenomenon in shared spaces and affect the safety and efficiency of such environments. To replicate group phenomena and systematically study their influence in shared spaces; realistic models of social groups and the integration of these models into shared space simulations are required. In this work, we focus on pedestrian groups and adopt an extended version of the social force model in conjunction with a game-theoretic model to simulate their movements. The novelty of our paper is in the modeling of interactions between social groups and vehicles. We validate our model by simulating
scenarios involving interaction between social groups and also group-to-vehicle interaction.

\end{abstract}
\section{Introduction}
Over the past years, the shared space design principle has been studied as an alternative to traditional regulated traffic designs. In shared spaces, different types of road users, e.g., pedestrians (often in groups\cite{cheng2018modeling}), cyclists and vehicles coexist with little or no explicit traffic regulations.
Therefore, road users need to interact with each other more frequently to negotiate their trajectories and avoid conflicts based on social protocols and informal rules. According to ~\cite{gettman2003surrogate}, we define conflict as “an observable situation in which two or more road users approach each other in time and space to such an extent that there is a risk of collision if their movements remain unchanged.”

Studying and understanding heterogeneous road users' movement behaviors under different circumstances and providing realistic simulation models provide a good basis for analyzing traffic performance and safety of shared spaces.
Modeling mixed traffic interactions is challenging because of the diversity of user types and also of road users behavior.
Even in the same user class, each may have a different point of view according to their characteristics.

Pedestrian groups form a large part (70\%) of the crowd population\cite{moussaid2010walking}. 
However, previous works on shared space simulation \cite{johora2018modeling,schonauer2012modeling,anvari2015modelling} have been mostly ignoring social group phenomena. 
In \cite{rinke2017multi}, single pedestrian-to-pedestrian group and group-to-group interactions are considered. Works on crowd behavior modeling have paid considerable attention to group dynamics but in homogeneous environments \cite{vizzari2013adaptive,huang2018social,kremyzas2016towards}.
In \cite{vizzari2013adaptive}, G. Vizzari et al. proposed a Cellular Automata (CA) based model to describe two types of groups; namely, simple group (a small set of pedestrians) and structured group (a large set of pedestrians which can be structured into smaller subgroups), by considering the goal orientation and cohesion of group.
In \cite{huang2018social}, the classical Social Force Model (SFM) is extended to 
describe both intra-group and inter-group interactions of pedestrian groups.

A. Kremyzas et al. also proposed an extension of SFM in \cite{kremyzas2016towards}, called \textit{Social Groups and Navigation (SGN)} to simulate the behavior of small pedestrian groups. In addition to group coherence behavior, SGN can model the situation where a group splits to even smaller groups when needed (e.g. in the overcrowded area) and reforms to re-establish its coherence.  

None of these works has considered group-to-vehicle interaction which is challenging because of the heterogeneity in the behavior of different groups. In this paper, we take the first step to address this gap by extending the multiagent-based model described in \cite{johora2018modeling} to incorporate and model pedestrian groups behavior while interacting with vehicles, single pedestrians and other groups. 

\section{Problem Statements and Requirements} Understanding the movement behaviors of road users in shared spaces is important to model these environments.

To understand the typical interactions of pedestrians and vehicles, in~\cite{johora2018modeling}, we analyzed the data of a road-like shared space area in Hamburg \cite{rinke2017multi}. From our observation and based on the classification of road user behavior given in \cite{helbing1995social}, we proposed grouping these interactions into two categories, based on the complexity of interaction:
\begin{itemize}
    \item \textbf{Simple Interaction:} Direct mapping of road users perceptions to their actions.
     \begin{itemize}
         \item \textbf{Reactive Interaction:} Road users behave reactively without further thinking to avoid urgent or sudden conflicts. As an example, if a vehicle suddenly stops because a pedestrian suddenly appears in front of the vehicle, then the vehicle behind also needs to perform an emergency break to avoid a serious collision.
         \item \textbf{Car Following (Vehicle only):} Empirical observation states that although there is no defined lane for vehicles in shared spaces, they follow the vehicle in front as they are driving into an assumed lane~\cite{anvari2015modelling}.
     \end{itemize}
    \item \textbf{Complex Interaction:} Road users choose strategy among different alternative strategies.
    \begin{itemize}
        \item \textbf{Implicit Interaction:} Road users choose their best action by predicting others’ action. As a real-life example, when a pedestrian crosses a road with a high speed, an approaching vehicle might predict that the pedestrian will continue to walk, so the driver will decelerate to avoid collision.
        \item \textbf{Explicit Interaction:}  Road users also interact with others using hands or eye contact as a means of communication.
    \end{itemize}
\end{itemize}

In this paper, we analyze the properties of pedestrian groups and their interaction with others. The term 'Group' here does not mean a couple of random people that happen to walk close to each other. According to \cite{moussaid2010walking}, we define a group term as "it is not only referring to several proximate pedestrians that happen to walk close to each other, but to individuals who have social ties and intentionally walk together, such as friends or family members". We describe pedestrian groups by the following features:
\begin{itemize}
    \item \texttt{} \emph{\textbf{Size}}: The most frequent group sizes vary between two to four members, whereas groups comprising five or more members are rare according to \cite{moussaid2010walking}.
    \item \texttt{} \emph{\textbf{Goal Position}}: Normally all group members walk towards predefined a common destination.
    \item \texttt{} \emph{\textbf {Coherence}}: Coherence is an important criterion of groups; in other words, group members manage to stay together. However, if the group members split temporarily for any reason, the faster members would wait at a safe point until other members reach that point and the group becomes coherent again \cite{kamphuis2004finding}.
    \item \texttt{} \emph{\textbf{Speed}}: The average speed of pedestrians in a
group is dependent on the group size; bigger groups are slower compared to small groups and the relation between group size and average speed of group is linear \cite{moussaid2010walking}.
    \item \texttt{} \emph{\textbf{Clustering Option}}: We observe from real shared space data and also from the literature on group dynamics \cite{huang2018social,costa2010interpersonal} that social groups frequently split if the group size is greater than 3.
\end{itemize}

In case of group-to-group interaction, groups usually deviate from their current path instead of decelerating or accelerating, to avoid collision.
Interactions between a group of pedestrians and a vehicle are observed to happen typically in the following way: group members follow a temporary leader who decides on action (decelerate, accelerate or deviate) first or they split into subgroups (groups split even while walking \cite{huang2018social}), subgroups interact with the vehicle individually and re-form afterwards. The leader reacts in a similar way as a single person would, but by taking the group size and positions of the group members into consideration. 

\begin{figure}[htbp]\vspace*{4pt}
\vspace{-1.7em}
\centerline{\includegraphics[width=4.6in]{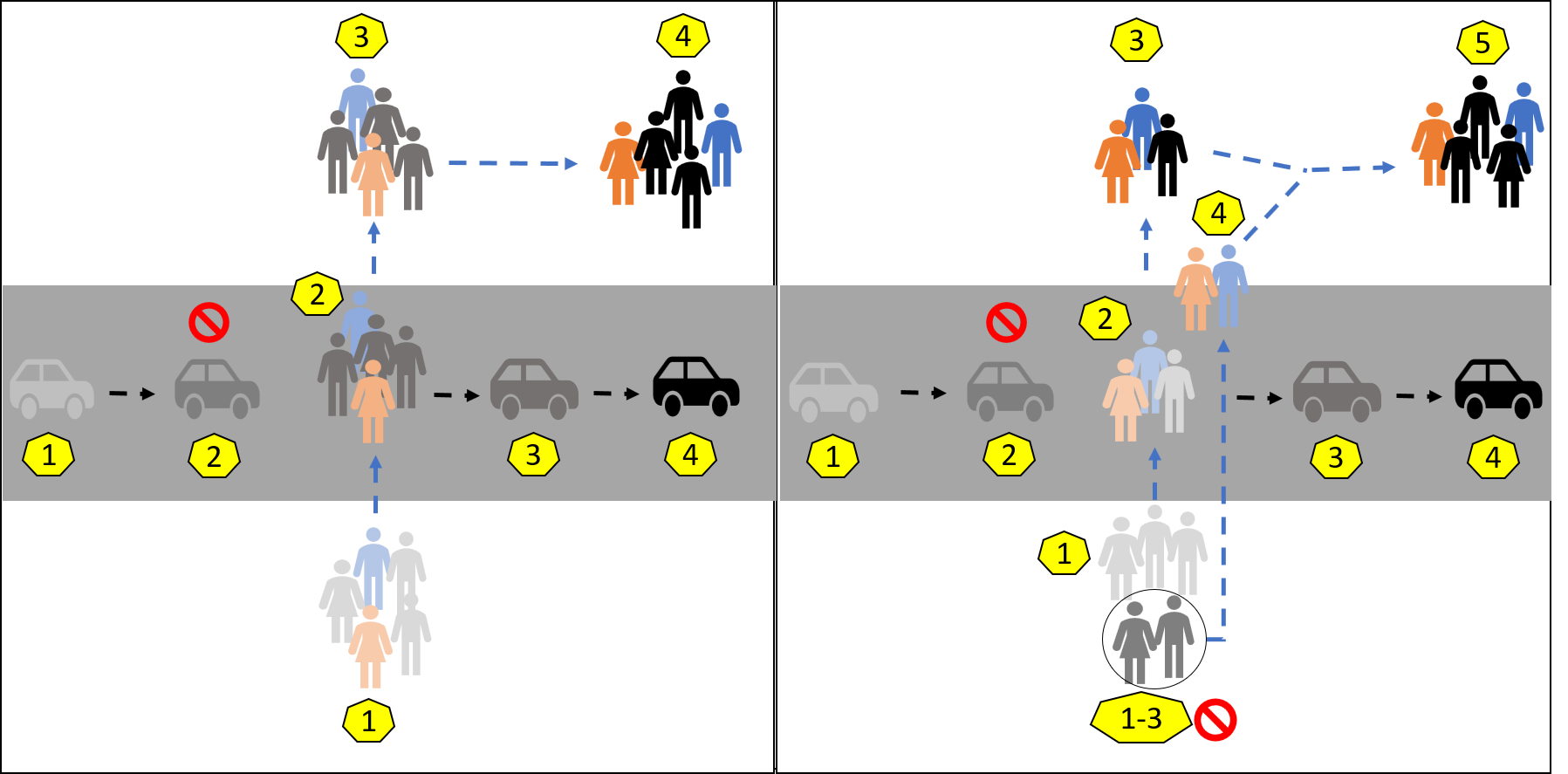}}
\caption{Non-clustered group (left) vs.~clustered group (right) interaction with a vehicle. Here, the numbers (1-5) represent the interaction steps and the lighter to darker shades of colors represent positions of car and pedestrians at different steps (lightest: first step and darkest: last step). The group leader and last member are visualized by blue and orange colors respectively.}\label{fig:cluster}
\vspace{-1.3em}
\end{figure}

In Figure~\ref{fig:cluster}, we depict an imaginary scenario to illustrate group-to-vehicle interaction behavior: where a car $Car_1$ (moving to the \emph{right} direction) interacts with a pedestrian group $G_1 = \{M_1,M_2,M_3,M_4,M_5\}$ (crossing a road from \emph{bottom to top} direction). Lets say, in this case, the outcome of their interaction is that $Car_1$ will wait and all or some members of $G_1$ will continue.  
If $G_1$ is a non-clustered group with one leader member who makes decision in any situation and all other members follow his/her decision, then the interaction between $G_1$ and $Car_1$ is executed as shown in the following steps: 

\begin{itemize}
	\item $1$: $Car_1$ starts decelerating to let the group cross and  $G_1$ starts crossing the road. 
	\item $2$: $Car_1$ eventually stops and  $G_1$ continues crossing the road.
	\item $3$: $G_1$ almost crossed the road by maintaining group coherence i.e. all group members move together as a single group without splitting into smaller subgroups. $Car_1$ starts moving as $G_1$ is at a safe distance.
	\item $4$: Both $Car_1$ and $G_1$ continue moving towards their destinations.
\end{itemize}

Whereas, if $G_1$ is a clustered group, then the execution of interaction between $G_1$ and $Car_1$ is performed in the following steps: 

\begin{itemize}
	\item $1$: $Car_1$ starts decelerating to let the group pass first. $G_1$ splits into two subgroups and the subgroup which contains the original group leader, $G_{1.1}$ starts crossing the road. Whereas, the other subgroup $G_{1.2}$ waits (other possible actions for $G_{1.2}$ are continue walking or deviate) for $Car_1$ to pass.
	
	\item $2$: $Car_1$ eventually stops, $G_{1.1}$ continues crossing the road and $G_{1.2}$ waits.
	
	\item $3$: $G_{1.1}$ crosses the road and waits for $G_{1.2}$ to re-form the group, and $Car_1$ starts moving as both subgroups are at safe distances.
	
	\item $4$: $Car_1$ continues driving, $G_{1.1}$ waits for $G_{1.2}$ and $G_{1.2}$ starts crossing the road.
	\item  $5$: $G_{1.1}$ and $G_{1.2}$ meet, re-form the group $G_1$ to reestablish their group coherence and then $G_1$ continues moving towards its destination.
\end{itemize}

We model the intra-group interaction and interaction of pedestrian group with single pedestrian or other groups (assumed to be similar to pedestrian-to-pedestrian interaction) as simple interaction, whereas we model the pedestrian group-to-vehicle interaction both as simple and complex interactions. The \textbf{simple interactions} can be modeled using force-based \cite{helbing1995social} or cellular automata models \cite{nagel1992cellular}. However, decision-theoretic models such as multinomial logit model \cite{helbing1995social} or game theoretic model \cite{schonauer2017microscopic,kita1999merging,lutteken2016using,bjornskau2017zebra} are more efficient for representing \textbf{complex interactions} \cite{helbing1995social}. In a logit model, a decision maker makes a decision regardless of others' decision based on present data \cite{pascucci2018should}, whereas, in a non-cooperative game-theoretic model, each decision maker includes predictions of other players' decisions in its own decisions. 

We note that in this paper we only consider single pedestrian, groups of pedestrians and vehicles as road users; modeling \textit{explicit} interaction is a topic of future work.

\section{Multiagent-Based Simulation Model}
\begin{figure}[htbp]\vspace*{4pt}
\centerline{\includegraphics[width=4.7in]{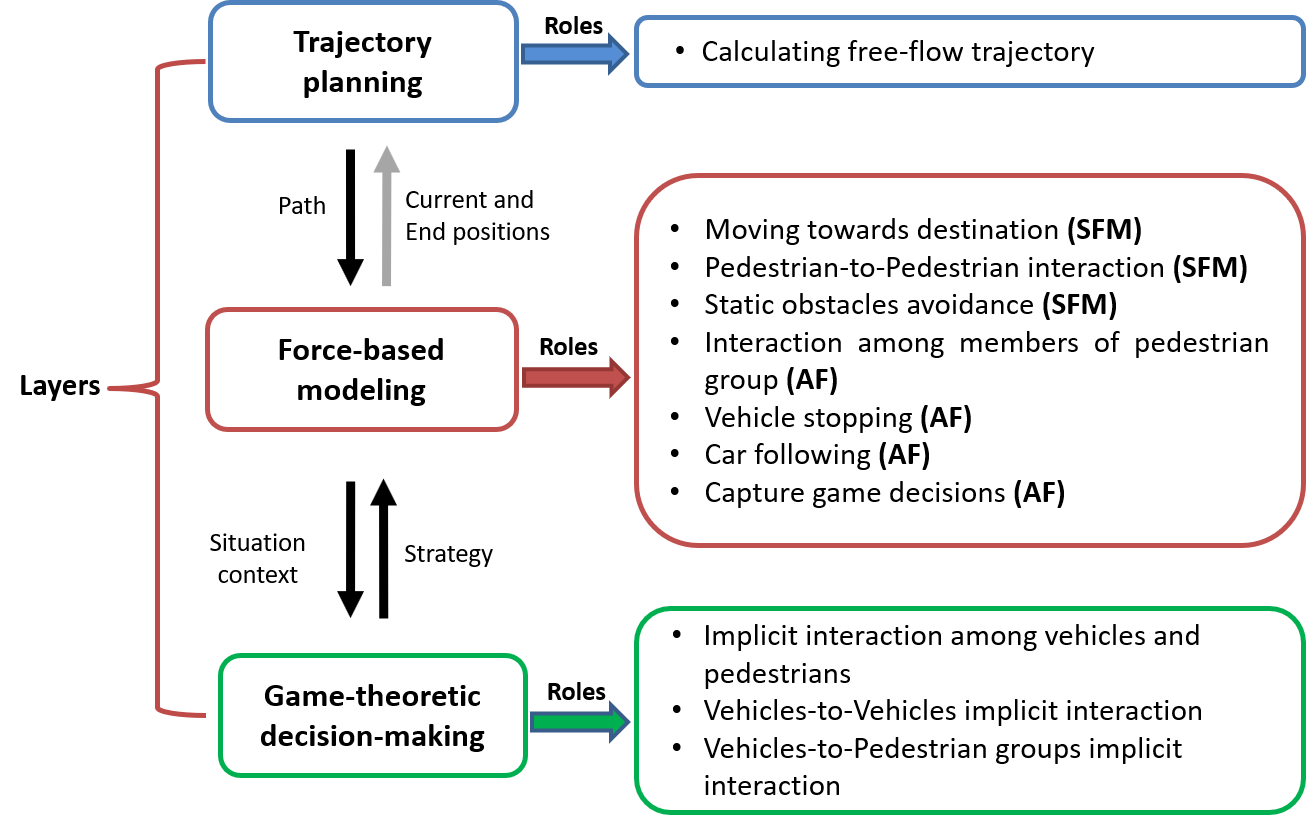}}
\caption{Conceptual model of road users' motion behaviors. AF means additional force.}
\label{fig:overallmodel}
\vspace{-1.3em}
\end{figure}

We integrate pedestrian group dynamics to our existing simulation model \cite{johora2018modeling} which comprises of three modules: trajectory planning, force-based modeling, and game-theoretic decision-making as visualized by Figure~\ref{fig:overallmodel}.
\subsection{Trajectory Planning Layer}
The trajectory planning module computes free-flow trajectories of each road user by considering
static obstacles such as the buildings boundaries or trees in the shared space environment. We convert our simulation environment into a visibility graph by connecting outline vertices of the obstacles, to perform the A* trajectory planning algorithm~\cite{moussaid2010walking}. We only connect two vertices if they are visible to each other and their resulting edge does not collide with any outline of any obstacle~\cite{koefoed2012representations}. We also add the origin and destination points of each road user to the graph. After planning the trajectories of road users, we adjust the position of their inner path vertices to capture the fact that humans tend to keep some distance from obstacles.
\vspace{-0.5em}
\subsection{Force-Based Modeling Layer}
We use the social force model (SFM), introduced by Helbing et al.~\cite{helbing1995social}, to model simple interactions of road users. 
In the classical SFM, the movement of a pedestrian is controlled by a set of simple forces, which are represented in equation~\ref{eq:CSFM}. These forces reflect the inner motivation of a pedestrian to perform certain actions based on the circumstances.

\begin{equation}
 \frac {d{\overrightarrow{v}_a}}{dt} :=	\overrightarrow{f}_{\alpha}^o + \Sigma\overrightarrow{f}_{\alpha B} + \Sigma\overrightarrow{f}_{\alpha \beta}
 \label{eq:CSFM}
\end{equation}

We apply the classical SFM to capture the driving force of road users towards their destination ($\overrightarrow{f}_{\alpha}^o$), their repulsive force towards static obstacles ($\overrightarrow{f}_{\alpha B}$) and towards other pedestrians ($\overrightarrow{f}_{\alpha\beta}$). We extend SFM to model car following interaction among vehicles ($\overrightarrow{I}_{following}$) and reactive interaction ($\overrightarrow{I}_{stopping}$) of vehicles towards pedestrians by decelerating to allow pedestrians to pass first. $\overrightarrow{I}_{stopping}$ only occurs if pedestrian(s) or pedestrian group(s) has already started moving in front or nearby to the vehicle. The details of these forces is in \cite{johora2018modeling}.

We select the Social Groups and Navigation model (SGN) of \cite{kremyzas2016towards} and the model of Moussaid et al.~\cite{moussaid2011simple} to capture the intra-group behavior of pedestrian groups. Both these models are extensions of SFM \cite{helbing1995social}. 
To model vehicle-to-pedestrian groups interaction, we extend these both models, see Subsection~\ref{subsec:interactionhandling}.

\begin{figure}[htbp]\vspace*{4pt}
\vspace{-1em}
\centerline{\includegraphics[width=4in]{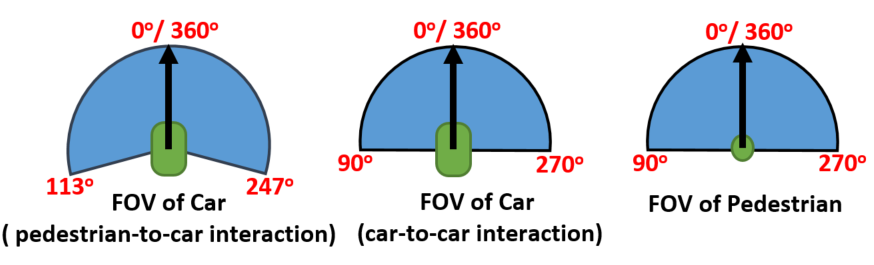}}
\caption{Road users field of view.}
\label{fig:FOV}
\vspace{-0.5em}
\end{figure}

The group force term $\overrightarrow{f}_{group}$ defines the interaction among group members. It combines two forces, namely, visibility force $\overrightarrow{f}_{vis}$ and attraction force $\overrightarrow{f}_{att}$ \footnote{vector + vector = vector; point + vector = point},
to help group members to stay connected and maintain the group structure. $\overrightarrow{f}_{vis}$ stands for the desire of a pedestrian to keep his/her group members within his/her field of view (FOV, see Figure~\ref{fig:FOV}) and $\overrightarrow{f}_{att}$ attracts any group member (except the leader) to the centroid of the group, when the member exceeds a calculated threshold value $d$ unless the respective member reaches his/her goal and as a result his/her desired velocity $V_{desired}$ becomes zero.

\begin{equation}
\label{eq:group}
\overrightarrow{f}_{group} =  \overrightarrow{f}_{vis} + \overrightarrow{f}_{att}
\end{equation}

\begin{equation}
\overrightarrow{f}_{vis} =  S_{vis} * \theta  * \overrightarrow{V}_{desired} 
\end{equation}

\begin{equation}
\overrightarrow{f}_{att}=\begin{cases}
S_{att} *{ \overrightarrow{n}( A_{ij},C_i) } , & \text{if $dist( A_{ij},C_i)\geq d \: and \: \overrightarrow{V}_{desired} \neq 0$}\\
0, & \text{otherwise}.
\end{cases}
\end{equation}

Here, $S_{vis}$ and $S_{att}$ are global strength parameters, $\theta$ is the minimum angle between every two members in a pedestrian group $G_i$ that they should maintain to stay within each other's FOV, $\overrightarrow{V}_{desired}$ denotes the desired velocity of any member $A_{ij}$ of $G_i$, and $\overrightarrow{n}( A_{ij},C_i)$ represents the distance between $A_{ij}$ and the centroid of the group ($C_i$), normalized in unit length. The $C_i$ is defined as follows: 
\begin{equation}
C_i = \frac{1}{|G_i|} \sum_{1\leq j \leq|G_i|} x_{ij} 
\end{equation}
Here, $|G_i|$ denotes the total number of members in $G_i$ and $x_{ij}$ depicts the position of any member $A_{ij}$ of $G_i$.

Execution of the game module decisions ($\overrightarrow{I}_{game}$) are also handled in this module, for example, if the result of a game-theoretic interaction between a vehicle driver and a group of pedestrians is that the groups can continue and the vehicle should wait (see also Subsection~\ref{subsec:game}), the respective actions of all users will be executed in this module. In this paper, we consider the continue, decelerate and deviate (for single or group of pedestrians only) as feasible strategies for road users and model these strategies as follows: 

\begin{itemize}
	\item Continue: Any pedestrian $\alpha$ crosses vehicle $\beta$ from the point $p_\alpha = {x}_\beta(t) + S_{c} * \overrightarrow{e}_\beta$, if  $line(x_\alpha(t), E_\alpha)$ intersects $line({x}_\beta(t) + S_{c} * \overrightarrow{e}_\beta, {x}_\beta(t) - \frac{S_c}{2} * \overrightarrow{e}_\beta)$, otherwise free-flow movement is continued. For vehicles, they continue their free-flow movement without any deviation. Here, scaling factor which depends on vehicle's speed is denoted by $S_c$, $\overrightarrow{e}$ is the direction vector, $x(t)$ and $E$ represent current and goal positions respectively.
	\item Decelerate: Road users decelerate and in the end stop (if necessary). For pedestrians, $newSpeed_{\alpha} = \frac{currentSpeed_\alpha}{2}$ and in case of vehicles, $newSpeed_{\beta} = currentSpeed_{\beta} - decelerationRate$.\newline
	\\
	Here, 
	$decelerationRate = \begin{cases}
    \frac{currentSpeed_{\beta}}{2}, & \text{if } distance(\alpha, \beta) \leq D_{min}, \newline
    \\
    \frac{currentSpeed_{\beta}^2}{distance(\alpha, \beta) - D_{min}}, & \text{otherwise}.
   \end{cases}$ 
   \newline
   $D_{min}$ is the critical spatial distance.
    \item Deviate: A pedestrian $\alpha$ passes a vehicle $\beta$ from behind from a position $p_\alpha = {x}_\beta(t) - S_d*\overrightarrow{e}_\beta(t)$ and after that $\alpha$ resumes moving towards its original destination. However, as long as $\beta$ stays in range of the field of view (FOV) of $\alpha$, $\alpha$ will keep moving towards $p_\alpha$. Here, $S_d$ is a scaling factor.
    The value of $S_c$, $D_{min}$, and $S_d$ need to be calibrated.
\end{itemize}

The overall resulting behaviors of pedestrians and vehicles are presented in equation~\ref{eq:PedSFM} and equation~\ref{eq:CarSFM} respectively.

\begin{equation}
 \frac {d{\overrightarrow{v}_a}}{dt} :=	\Big(\overrightarrow{f}_{\alpha}^o + \Sigma\overrightarrow{f}_{\alpha B} + \Sigma\overrightarrow{f}_{\alpha \beta} + \overrightarrow{f}_{group}\Big) \hspace{0.1cm} or \overrightarrow{I}_{game}
 \label{eq:PedSFM}
\end{equation}
\begin{equation}
 \frac {d{\overrightarrow{v}_a}}{dt} :=	\overrightarrow{f}_{\alpha}^o \hspace{0.1cm} or \overrightarrow{I}_{following} \hspace{0.1cm} or \overrightarrow{I}_{game} \hspace{0.1cm} or \overrightarrow{I}_{stopping}
 \label{eq:CarSFM}
\end{equation}
The game-theoretic decision has priority over the decision of this module for both user types, except for $\overrightarrow{I}_{stopping}$
of vehicles. 

\subsection{Game-Theoretic Decision Layer}
\label{subsec:game}
The game-theoretic decision module handles the implicit interactions between two or more road users. We use a sequential leader-follower game called Stackelberg game to handle these interactions, in which the leader player acts first and the followers react based on the leader’s action to maximize their utility \cite{schonauer2017microscopic}. We set the number of leaders to one and there can be one or more followers for any individual game. The vehicle (faster agent) is chosen as the leader in case of pedestrian(s)-to-vehicle and vehicle-to-group interactions, whereas in case of pedestrian(s)-to-multiple vehicles and vehicle-to-vehicle interactions, we pick one of these vehicles as the leader randomly. Only one game is played for each implicit interaction and the games are independent of each other. 

\begin{figure}[htbp]
	\centering
	\vspace{-1em}
	\subfloat[]{\includegraphics[width=2.25in,height=1.22in]{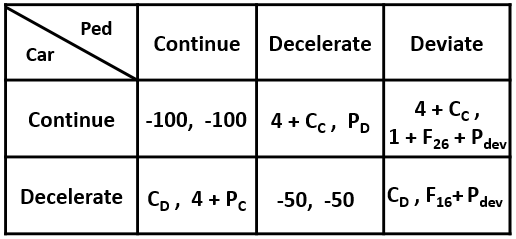}}
	\hfil
	\subfloat[]{\includegraphics[width=2.25in,height=1.22in]{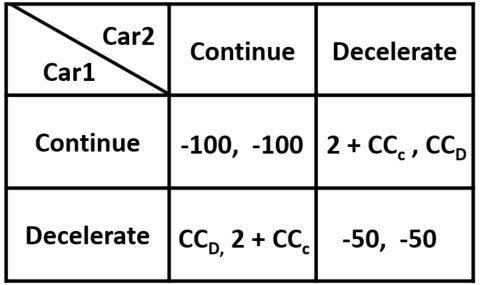}}
		\hfil
	\subfloat[]{\includegraphics[width=3in]{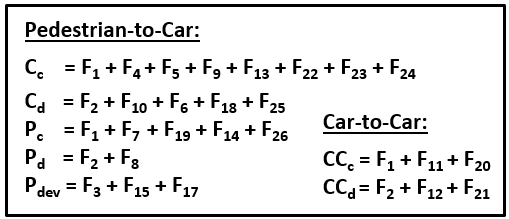}}
	\caption{The complete payoff matrices of pedestrian/group-to-vehicle and
	vehicle-to-vehicle interactions with all considered actions. (a) Pedestrian/group-to-vehicle Interaction (b) Vehicle-to-vehicle Interaction (c) Impacts of situation dynamics}
    \vspace{-0.9em}
    \label{fig:payoff}
\end{figure}
As projected by Figure~\ref{fig:payoff}, to calculate the payoff matrix of the game, firstly, we ordinary value all actions of the players from -100 to 4 (here, positive values are preferred outcome) with the assumption that reaching their destination safely and quickly is their main preference. Secondly, to capture courtesy behavior and situation dynamics, eleven relevant observable factors are taken into account. Among these factors, two are especially for groups-to-vehicle interaction. These factors are defined in the following as Boolean (1,0) variables $x_1 \ldots x_{11}$ and used to calculate a set of parameters $F_1 \ldots F_{26}$, which are impacts of these factors. Let $\alpha$ be a road user who interacts with another user $\beta$:
\vspace{-0.75em}
\begin{itemize}
	\item $x_1$: has value $1$, if current speed of $\beta$, $S_{current}$ $<$ $S_{normal}$. This factor determines the value of $F_1 \hspace{0.05cm}to\hspace{0.05cm} F_3$.
	\item $x_2$: has value $1$ if number of active interactions $<$ N. This factor decides the value of $F_4$.
	\item $x_3$: has value $1$ if $\alpha$ already stopping to give way to another user $\beta'$. This decides the value of $F_5 \hspace{0.05cm}to\hspace{0.05cm} F_8$ and $F_{16}$.
	\item $x_4$: has value $1$ if $\alpha$ is a car driver following another car $\beta'$. This factor determines the value of $F_9 \hspace{0.05cm}to\hspace{0.05cm} F_{12}$.
	\item $x_5$: has value $1$ if path deviation does not result long detour for $\alpha$ (pedestrian): ($\theta_{\overrightarrow{e}_\beta\hat{n}_{\alpha\beta}}$ $>$ 58\degree\hspace{0.05cm} \textbf{and} $\theta_{\overrightarrow{e}_\beta\hat{n}_{\alpha\beta}}$ $\leq$ 113\degree) \textbf{or} ($\theta_{\overrightarrow{e}_\beta\hat{n}_{\alpha\beta}}$ $\geq$ 247\degree\hspace{0.05cm} \textbf{and} $\theta_{\overrightarrow{e}_\beta\hat{n}_{\alpha\beta}}$ $<$ 302\degree). This factor decides the value of $F_{13}\hspace{0.05cm}to\hspace{0.05cm} F_{15}$.
	\item $x_6$: has value $1$ if $\alpha$ is a car driver followed by another car $\beta'$. This factor determines the value of $F_{17}$.
	\item $x_7$: has value $1$ if distance($\alpha$, $\beta$) $<$ M, then $\alpha$ (car) is unable to stop. This factor determines the value of $F_{18} \hspace{0.05cm}to\hspace{0.05cm} F_{19}$.
	\item $x_8$: has value $1$ if $\alpha$ as a car is in a roundabout. This factor determines the value of $F_{20}$ and $F_{21}$.
	\item $x_9$: to consider uncertainty in driving behavior, the value of $F_{22}$ generates randomly.
	\item $x_{10}$: has value 1 if the leader $\alpha$ of a group is in waiting state. This factor determines the value of $F_{23}$.
	\item $x_{11}$: has value 1 if a pedestrian $\alpha$ is in a group. This factor determines the value of $F_{24}$, $F_{25}$ and $F_{26}$.
\end{itemize}
The same variable $x_i$ is often used to determine multiple parameters e.g. $x_1$ calculates the value of $F_1$, $F_2$ and $F_3$, because, the same value of $x_i$ has different influences on different strategies of road users. As an example, if the speed of a pedestrian is less than her average speed, then $x_1$ has positive influence on strategy decelerate ($F_2$) and negative influence on continue ($F_1$). To analyze the behavior of our model, we perform a sensitivity analysis of the parameters $F_1 \hspace{0.05cm}to\hspace{0.05cm} F_{26}$, N and M on a certain amount of interaction scenarios. However, as we have not calibrate these parameters with a significant amount of real interaction scenarios, we do not present these parameters values in this paper. As part of our future work, we will focus on automated calibration and validation of these parameters. More details regarding these modules, game solving, interaction categorization and modeling can be found in~\cite{johora2018modeling}.

\subsection{Interaction Handling of Pedestrian Groups:}
\label{subsec:interactionhandling}
We have assigned some properties for each group which are essential to define, model and simulate the group behaviors: a specific ID, a leader $L_{ij}$, a boundary member $Lm_{ij}$ (the one with the largest distance to the leader) which is changeable during simulation \cite{kremyzas2016towards} and group splitting behavior; if the size $S$ of any group $G$ is as such that S $\geq$ 3, then G splits into subgroups ($G_1$,..,$G_n$) based on the probability $P$ defined as $P_{base} + (S - 3) * \alpha$. The calibration of the values of $P_{base}$ and $\alpha$ is part of our future research. 

\begin{figure}[H]\vspace*{4.5pt}
\centerline{\includegraphics[width=4.5in]{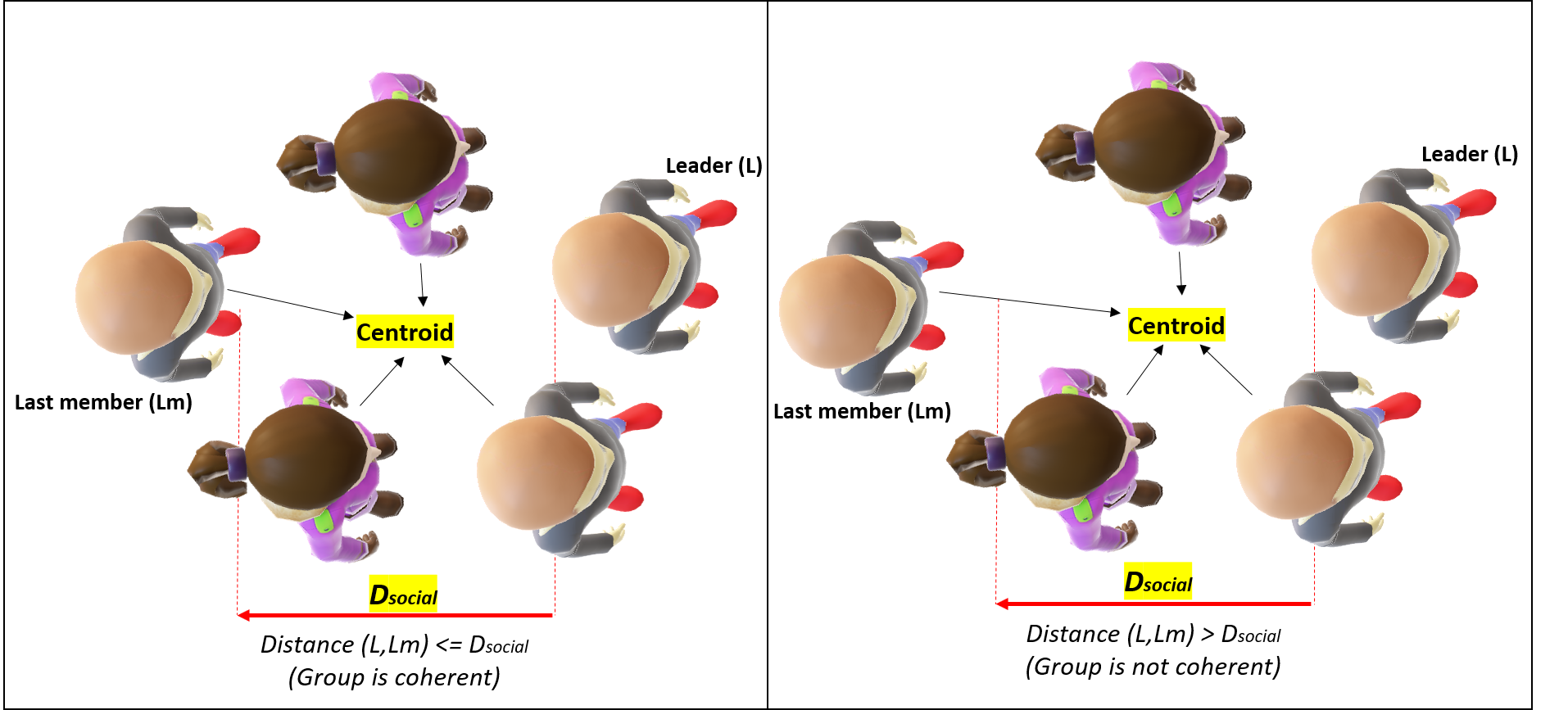}}
\caption{Coherent group (left) vs.~non-coherent group (right)}
\label{fig:coherentvsnoncoherent}
\end{figure}

In our model, we choose $L_{ij}$ in three different methods: namely, the nearest member to the competitive vehicle, to the group destination or to the road boarders. A group is called a coherent group if the distance between the last member $Lm_{ij}$ and leader $L_{ij}$ does not exceed a specific threshold $d_{social}$ \cite{kremyzas2016towards}. Figure~\ref{fig:coherentvsnoncoherent} shows a coherent and non-coherent group structure example.
\begin{equation}
 dist(x_{ij}, x_{ij'}) <= d_{social} 
 \end{equation}
\subsubsection{Intra-Group Interaction:}
\label{group_interaction}
\begin{figure}[htbp]\vspace*{4pt}
\vspace{-1.3em}
\centerline{\includegraphics[width=3.7in,fbox]{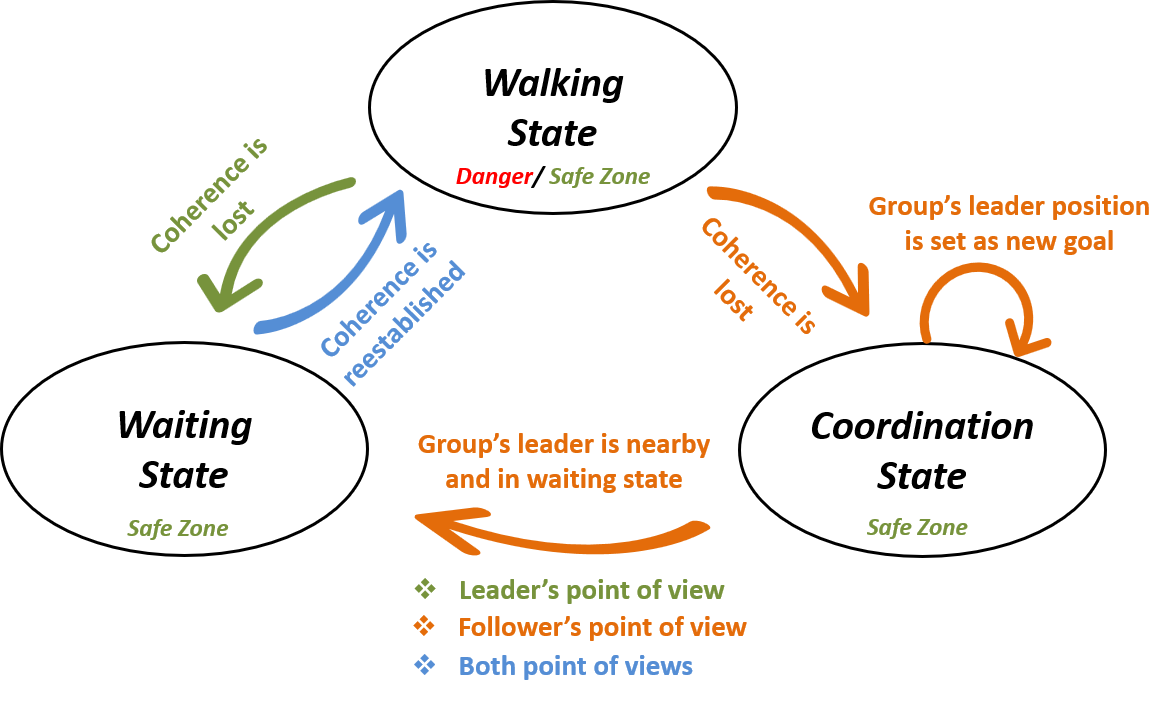}}
\caption{The movement states of pedestrian groups}
\label{fig:pedgroupstate}
\vspace{-1.3em}
\end{figure}

We model the interconnection of group members in three states i.e. walking, waiting and coordination states, illustrated in Figure~\ref{fig:pedgroupstate}. 

\begin{figure}[H]
\centerline{\includegraphics[width=3in]{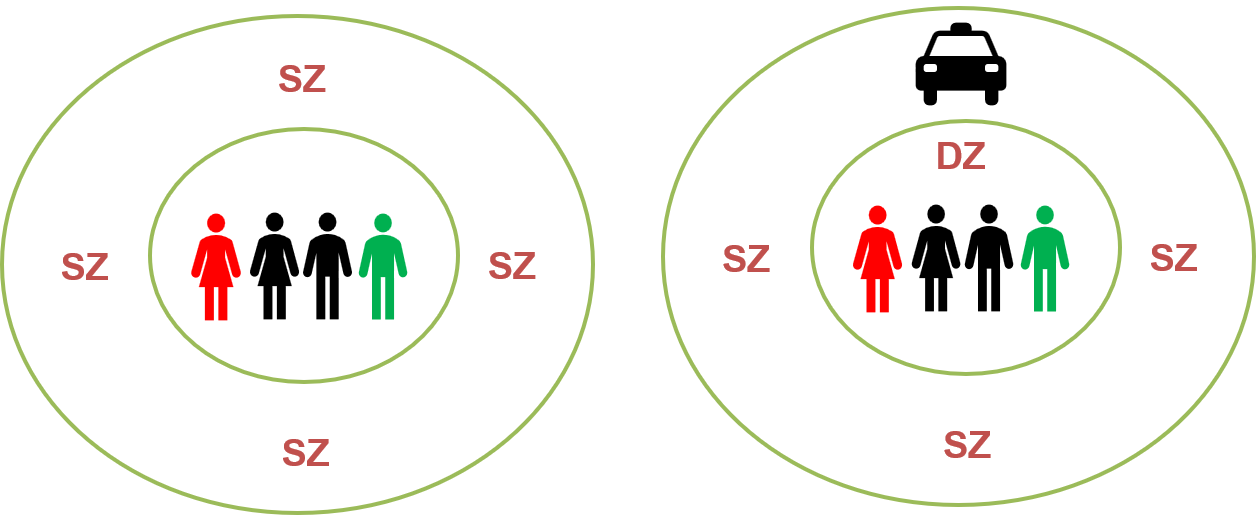}}
\caption{Safe zone SZ (left) vs.~danger zone DZ (right)}
\label{fig:zones}
\end{figure}

At the beginning of simulation, all group members are in the walking state, in which they walk together using equation~\ref{eq:PedSFM}. When the coherence of the respective group $G$ gets lost, group leader $L_{ij}$ switches to the waiting state and waits for the other members, whereas the other members switch to coordination state. In coordination state, the current position of $L_{ij}$ is set as a temporary goal of other members and they stay in this state until they reach to the leader's position. When every member reaches his/her leader, the coherence of $G$ is reestablished and therefore all group members return back to the walking state. However, group members can stay at waiting or coordination states only if they are in safe zone (a place which is safe to wait or coordinate). Pedestrian zones, and mixed zones with no competitive users within any particular group member's field of view can be examples of safe zones. Whereas, overcrowded pedestrian or mixed zones are defined as danger zone, see Figure~\ref{fig:zones}. 
\newline
\begin{equation}
f_{group}=\begin{cases}
f_{vis} + f_{att}, & \text{if $Safe\: zone$ and not in waiting state }\\
0, & \text{if $Danger\:zone$}.
\end{cases}
\end{equation}

\subsubsection{Pedestrian Group-to-Vehicle Interaction:}
\label{g_to_v}
\begin{figure}[htbp]\vspace*{4pt}
\vspace{-0.8em}
\centerline{\includegraphics[width=4.1in,fbox]{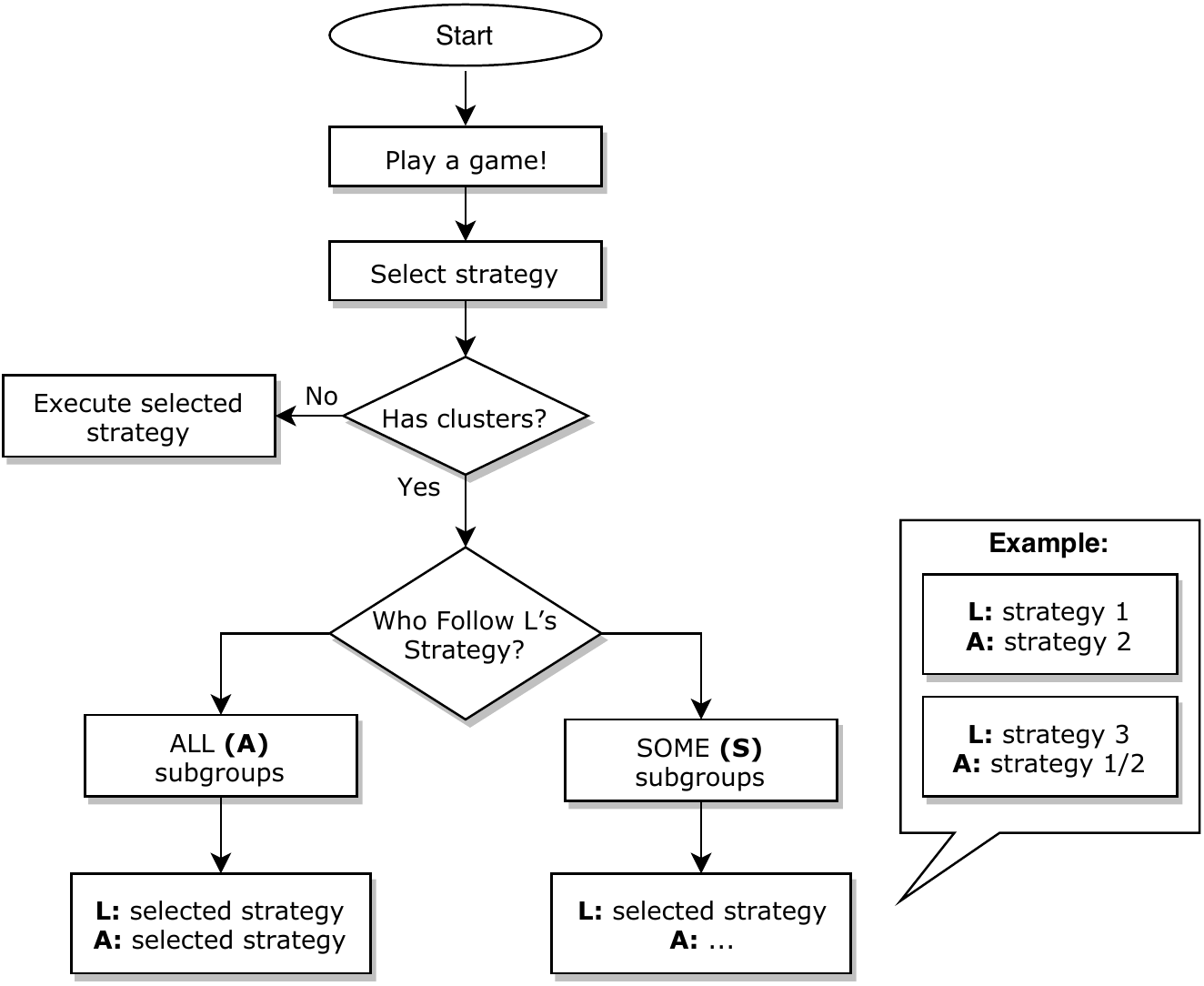}}
\caption{Execution steps of vehicle-to-pedestrian groups interaction. Here, strategy 1, 2 and 3 represent decelerate, deviate and accelerate, respectively.}
\label{fig:stepsofgameplaying}
\vspace{-0.9em}
\end{figure}

Interaction between any pedestrian group (G) and vehicle (V) is handled by considering the steps in Figure~\ref{fig:stepsofgameplaying}.
\begin{itemize}
\item First V and L decide on their strategy among different alternatives (decelerate, accelerate, deviate ) by playing a Stackelberg game, see Subsection~\ref{subsec:game}. 
\item If G splits, all subgroups of G either follow the strategy chosen by the leader. 
\item Or all members of the subgroup of G ($G_i$) which L belongs to, follow the strategy of L and all other subgroups can perform one of the following:
\begin{itemize}
    \item all subgroups can follow the strategy of L.
    \item or if V and L decide to decelerate and accelerate respectively, then other subgroups can either decelerate or deviate or some subgroups decelerate and others deviate.
    \item or if V accelerates and L decelerates, then all other subgroups deviate.
\end{itemize}
\end{itemize}

\section{Evaluation}
To validate our model, we extract and simulate 20 scenarios from a data set of a shared-space in Hamburg \cite{pascucci2015modeling}, which involve pedestrian groups-to-vehicle conflicts and visualize the difference in real and simulated trajectories and speed of all involved road users. We select one 
scenario among all those scenarios to illustrate
elaborately vehicle-to-group interaction and create two example scenarios, one to capture the interaction between big pedestrian groups and another to show the clustering feature of group. For the first scenario, we compared the real and simulated behaviors and trajectories of the involved road users. However, we could not compare the performance of the last two scenarios as real data is currently missing.

The implementation of our model is performed in Lightjason, a Java-based BDI multi-agent
framework \cite{aschermann2016lightjason}. An Intel Core\texttrademark i5 processor with 16 GB RAM was used to perform all the simulation runs.

\begin{figure}[H]
\vspace{-1.3em}
\centering
\sbox{\measurebox}{%
  \begin{minipage}[b]{.3\textwidth}
  \subfloat
    []
    {\label{fig:figA}\includegraphics[width=\textwidth,height=5cm]{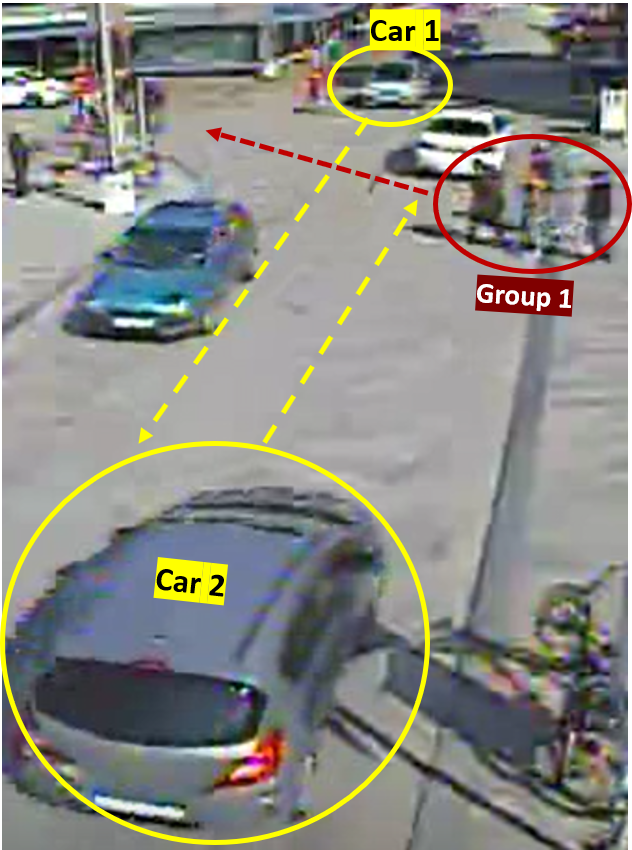}}
  \end{minipage}}
\usebox{\measurebox}\qquad
\begin{minipage}[b][\ht\measurebox][s]{.4\textwidth}
\centering
\subfloat
  []
  {\label{fig:figB}\includegraphics[width=\textwidth,height=2cm]{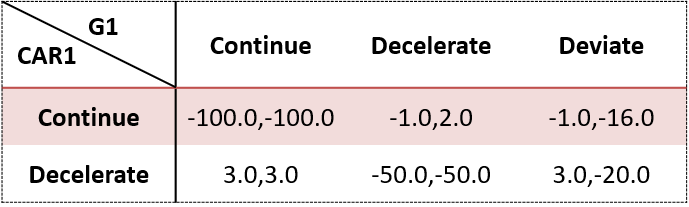}}

\vfill

\subfloat
  []
  {\label{fig:figC}\includegraphics[width=\textwidth,height=2cm]{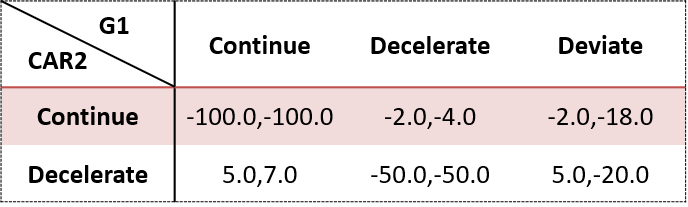}}
\end{minipage}
\begin{minipage}[b]{.78\textwidth}
  \subfloat
    []
    {\label{fig:figD}\includegraphics[width=\textwidth,height=8.2cm]{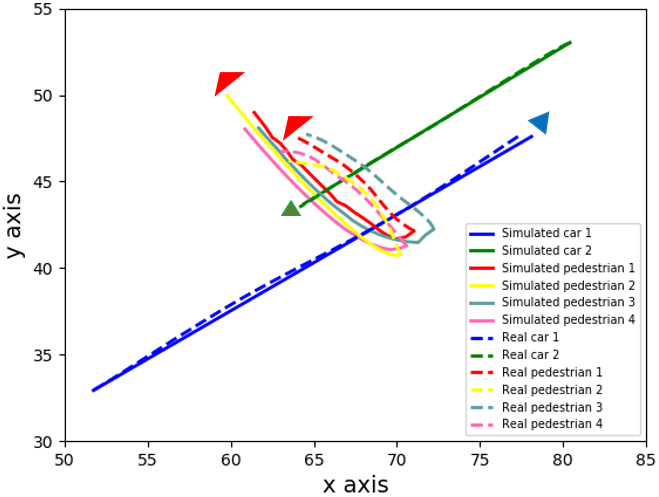}}
  \end{minipage}
\caption{Scenario 1: A pedestrian group to car interaction. (a) Real scenario snapshot (b) Game matrix: Car1 to Group1, (c) Game matrix: Car2 to Group1 (d) Comparison of the trajectories. The arrows indicate the direction of movement of the road users.}
\label{fig:realexp}
\end{figure}

\paragraph{\textbf{Scenario 1:}}
Here as shown in Figure \ref{fig:figA} an interaction between one pedestrian group (\emph{Group1}) of four members and two cars (\emph{Car1} and \emph{Car2}) is captured. In this scenario, both drivers decelerate to let the group cross the street. In our model, both these car-to-group interactions are handled using game-theoretic model. The game matrices in Figure \ref{fig:figB} and Figure \ref{fig:figC} reflect courtesy of the drivers towards the pedestrian group. 
In Figure~\ref{fig:figD}, the trajectories of both cars are very similar in simulation and real data, but the trajectories of group members in simulation are deviated from the real trajectories, even though the reacted behavior is same.

\begin{figure}[htbp]\vspace*{4pt}
\vspace{-1em}
\centerline{\includegraphics[width=4.1in]{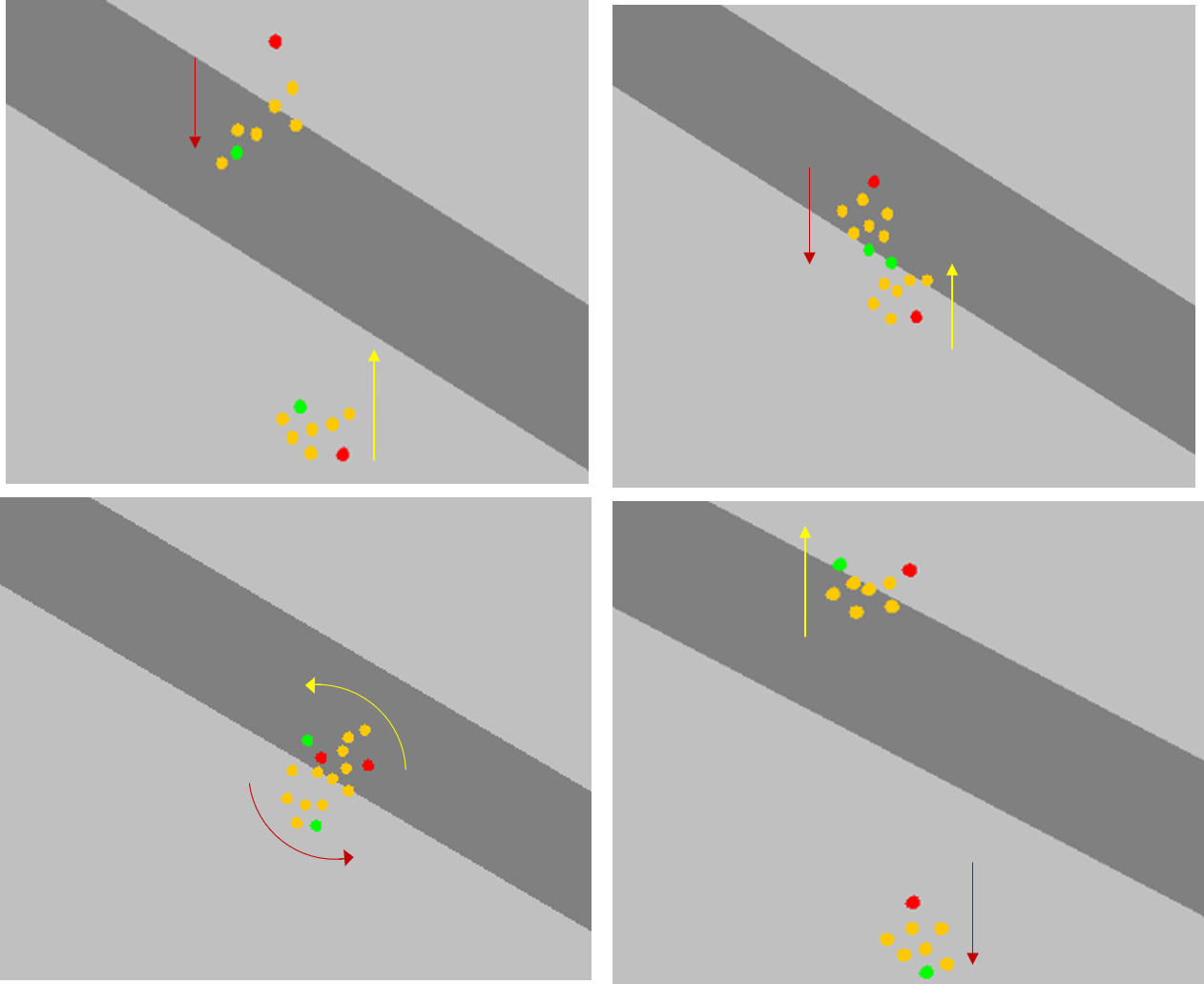}}
\caption{ Scenario 2: Non-clustered group to group interaction}
\label{fig:grouptogroup}
\vspace{.8em}
\end{figure}

\paragraph{\textbf{Scenario 2:}}
This scenario presents the interaction between two pedestrian groups of non-cluster structure. Each group consists of eight members.\emph{Group1} is moving to the southern direction, and \emph{Group2} is moving to the northern direction. As can be seen in Figure~\ref{fig:grouptogroup}, when the groups become very adjacent to each other, they slightly deviate to avoid collision with other. Our model handles this interaction as simple interaction.

\begin{figure}[H]
	\centering
	\subfloat[]{\label{fig:thirdexpmatrix}\includegraphics[width=2.4in,height=2cm]{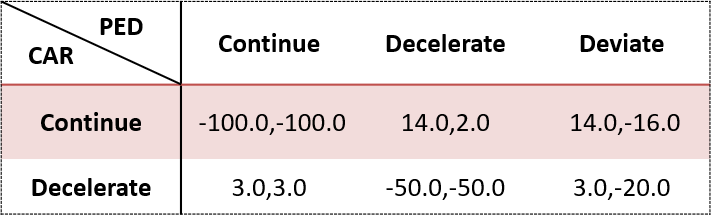}}
	\caption{Scenario 3: Clustered group to car interaction. (a) Simulation snapshots (b) Game matrix}
    \vspace{-0.75em}
    \label{fig:thirdScenario}
\end{figure}

\paragraph{\textbf{Scenario 3:}}
In Figure~\ref{fig:thirdScenario}, the interaction between a group $Group1$ and a car $Car1$ is represented. In this scenario, $Group1$ consists of six members and two clusters, where each cluster consist of three members. Since the leader $L$ of $Group1$ is in Safe Zone (car is still far away) and lost sight to the last member $Lm$, $L$ waits for his/her group members to reach $L$ to reestablish the group coherency before crossing the road. Afterwards, when $Car1$ and $Group1$ interact $Car1$ takes priority and continues to go because $L$ is in waiting state. The payoff matrix in Figure~\ref{fig:thirdexpmatrix} also reflects this factor.
\newline
To sum up, in all these scenarios, the pedestrian groups and cars suitably detect and classify the interaction and behave accordingly during simulation.

\begin{figure}[htbp]
	\centering
	\vspace{-1.3em}
	\subfloat[]{\includegraphics[width=2.37in,height=5cm]{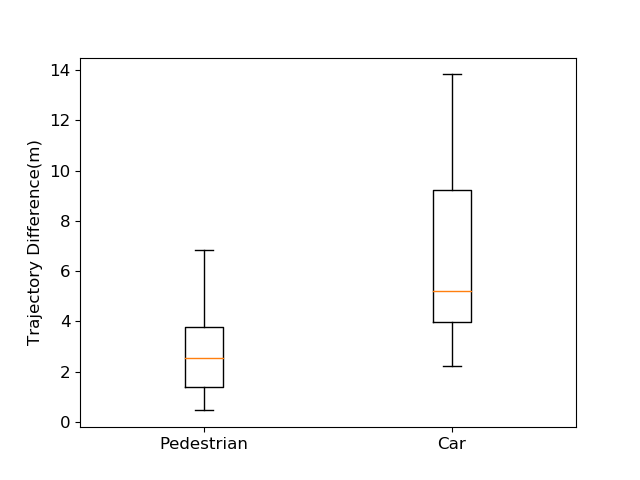}}
	\vspace{-0.3em}
	\subfloat[]{\includegraphics[width=2.37in,height=5cm]{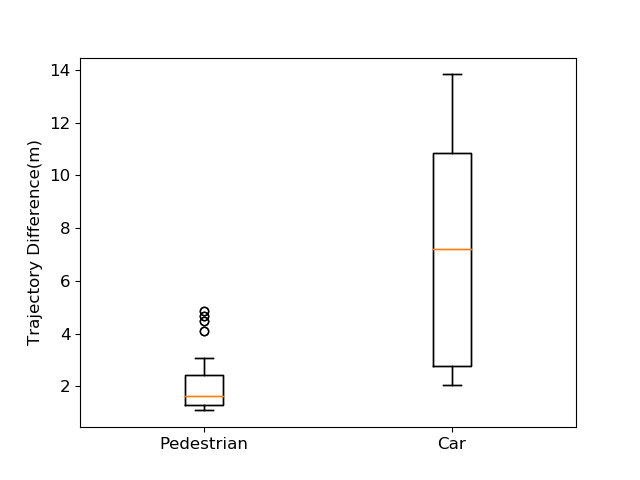}}
	\caption{Trajectory differences of real and simulated road users. (a) Without group model (b) With group model}
    \vspace{-0.75em}
    \label{fig:trajectorydev}
\end{figure}

\begin{figure}[H]
	\centering
	\vspace{-1.3em}
	\subfloat[]{\includegraphics[width=2.37in,height=5cm]{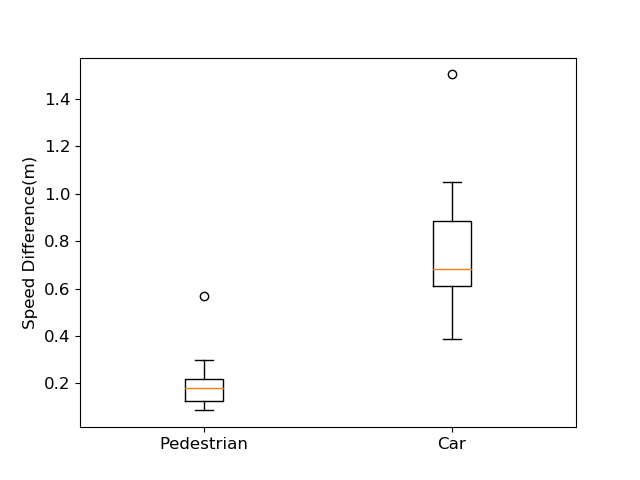}}
	\vspace{-0.3em}
	\subfloat[]{\includegraphics[width=2.37in,height=5cm]{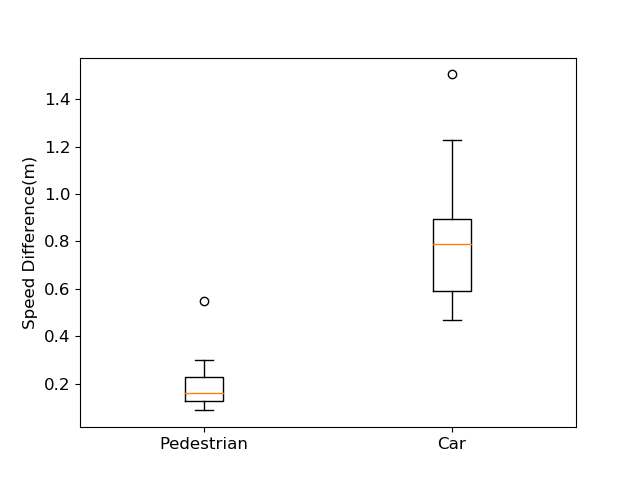}}
	\caption{Speed differences of real and simulated road users. (a) Without group model (b) With group model }
    \vspace{-0.75em}
    \label{fig:speedDev}
\end{figure}

Figure \ref{fig:trajectorydev} and Figure \ref{fig:speedDev} represent the average deviation of trajectories and speed of all road users of our selected 20 scenarios, before and after adding group dynamic to our model. The purpose of this comparison is to see the difference in interaction of car with arbitrary groups (e.g. a group of people waiting to cross a road) vs social groups (e.g. family members or socially related pedestrians walking together). 
The speed profiles of pedestrians and cars remain almost same after integrating pedestrian group behaviors to our model. Whereas, there is some improvement in terms of pedestrians trajectories (2.89 vs 2.21), but small weakening in cars trajectories (6.58 vs 7.0). Our analysis through simulation modeling says that, social groups take more time compared to arbitrary group to cross in front of the car, hence car has to wait longer causing this fore-mentioned weakening in car trajectory.

\begin{figure}[H]
	\centering
	\vspace{-1em}
	\subfloat[]{\includegraphics[width=4.1in]{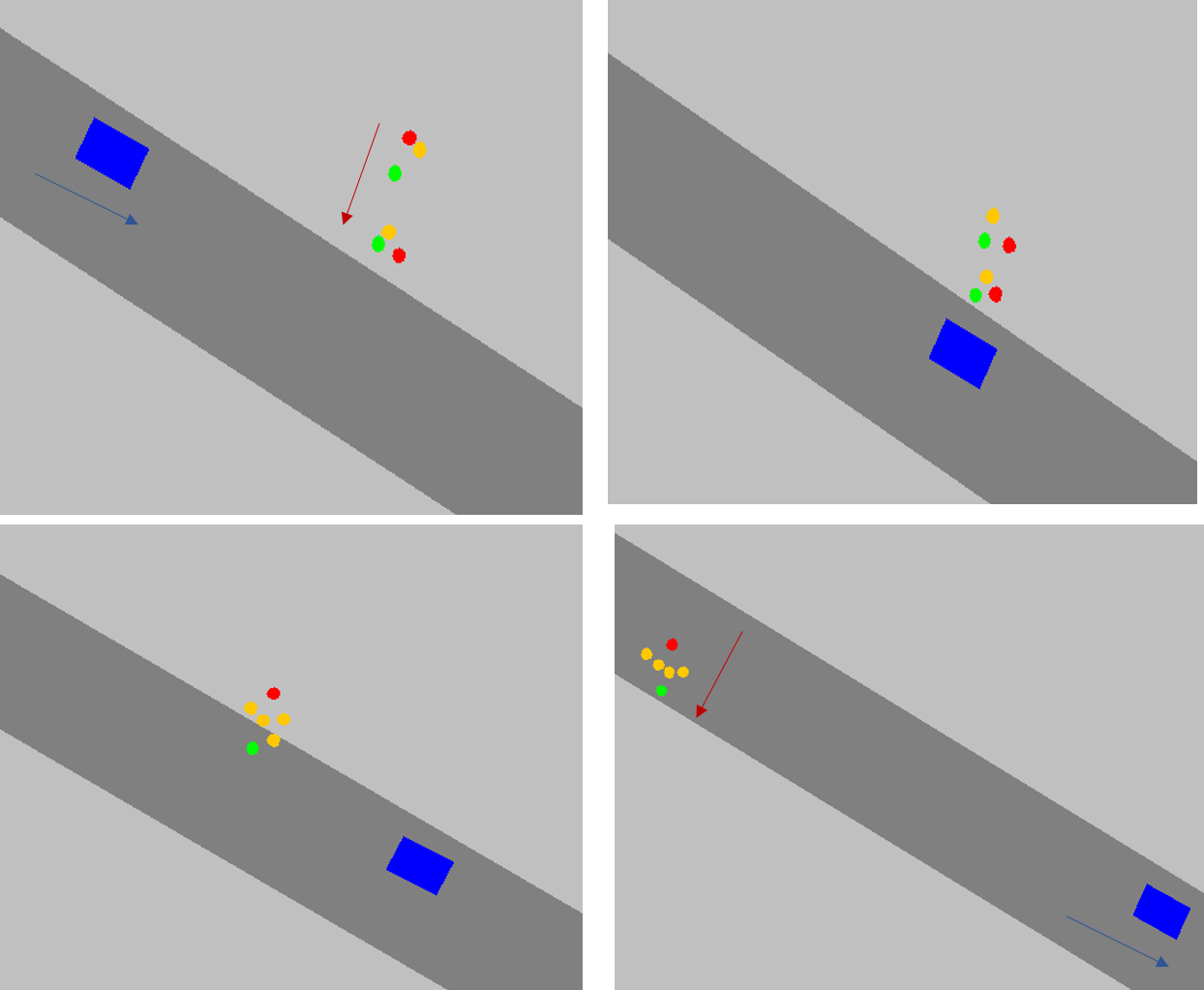}}
\end{figure}



\section{Conclusion and Future Work} 
In this paper, we integrate pedestrian group dynamic into our multiagent-based simulation model. Modeling of interactions between social groups and vehicles is our novel contribution. We analyze the interaction among pedestrian group members and also pedestrian group-to-vehicle interaction, classify them in terms of complexity and model these interactions using the social force model and Stackelberg games. First results support that our model has good potential to model both homogeneous (e.g. group-to-group) and heterogeneous (i.e. group-to-vehicle) interactions. We take a first attempt to investigate the difference in interaction of vehicle with arbitrary groups and social groups.

Our future research will be dedicated on calibrating and validating our model parameters, analysing the transferability of our model, investigating other influencing factors such as personal preferences, ages, and gender of road users, which might have impact on their decision making, and integrating explicit interaction among road users into our model.
Most importantly, we need to analyze and model larger scenarios with larger numbers of interactions between heterogeneous road users to examine the scalability of proposed interaction types and our simulation model.

\section{Acknowledgements}
This research is supported by the German Research Foundation (DFG) through the SocialCars Research Training Group (GRK 1931). We acknowledge the MODIS DFG project for providing datasets.

%
%
%
\bibliographystyle{splncs04}
\bibliography{mybibliography}

\end{document}